%% file: soi_cracow_sendai2015arXiv.tex
\documentclass[12pt]{article}
\usepackage{graphicx}
\usepackage{float}

\def\agh{AGH University of Science and Technology, Cracow, Poland}
\def\ifj{The Institute of Nuclear Physics PAN, Cracow, Poland}

\def\Title#1{\begin{center} {\Large #1 } \end{center}}
\def\Author#1{\begin{center}{ \sc #1} \end{center}}
\def\Address#1{\begin{center}{ \it #1} \end{center}}

\newenvironment{Abstract}{\begin{quotation}  }{\end{quotation}}
\newenvironment{Presented}{\begin{quotation} \begin{center} 
             PRESENTED AT\end{center}\bigskip 
      \begin{center}\begin{large}}{\end{large}\end{center} \end{quotation}}
\def\Acknowledgements{\bigskip  \bigskip \begin{center} \begin{large}
             \bf ACKNOWLEDGEMENTS \end{large}\end{center}}

\input econfmacros.tex
  
\usepackage{multicol}

\begin{document}
\begin{titlepage}

\vfill
\Title{Development of SOI pixel detector in Cracow }
\vfill
\Author{ Sz. Bugiel$^{*}$, R. Dasgupta$^{*}$, S. Glab$^{*}$, M. Idzik$^{*}$, J. Moron$^{*}$, P. Kapusta$^{\dagger}$, W. Kucewicz$^{*}$, M. Turala$^{\dagger}$}

\Address{$^{*}$ \agh \\
		 $^{\dagger}$ \ifj }
\vfill
\begin{Abstract}
This paper presents the design of a new monolithic Silicon-On-Insulator pixel sensor in $200~nm$ SOI CMOS technology. The main application of the proposed pixel detector is the spectroscopy, but it can also be used for the minimum ionizing particle (MIP) tracking in particle physics experiments. For this reason few different versions of pixel cells are developed: a source-follower based pixel for tracking, a low noise pixel with preamplifier for spectroscopy, and a self-triggering pixel for time and amplitude measurements. 
In addition the design of a Successive Approximation Register Analog-to-Digital Converter (SAR ADC) is also presented. A 10-bit SAR ADC is developed for spectroscopic measurements and a lower resolution 6-bit SAR ADC is integrated in the pixel matrix as a column ADC, for tracking applications. 
\end{Abstract}
\vfill
\begin{Presented}
International Workshop on SOI Pixel Detector, SOIPIX2015\\
Sendai, Japan,  June 3--6, 2015
\end{Presented}
\vfill
\end{titlepage}
\def\thefootnote{\fnsymbol{footnote}}
\setcounter{footnote}{0}

\section{Introduction}
Silicon-On-Insulator (SOI) technology process is known for about thirty years. Nowadays, it is becoming more and more popular and in many cases it successfully competes with the standard CMOS solutions. The SOI is very attractive for high speed and low power applications due to reduction of parasitic capacitances when compared with  the standard CMOS technology \cite{bulk}. 
Reducing the parasitic capacitances and leakage currents results in limiting the power dissipation and increasing the speed performance of the chip. Moreover, the elimination of leakage path to bulk allows the transistors to work over a very large temperature range from 4~K to 600~K \cite{arai2}.
The SOI is clearly a promising technology for the imaging applications because of a presence of the insulator - Burried Oxide Layer (BOX) - between the substrate and the thin layer of silicon with electronics\cite{Arai4}. Such separation allows a flexible  usage of the substrate as a radiation sensor  and the on-chip  electronics as the readout and processing circuitry.

	
\begin{figure}[ht]
\centering
\includegraphics[width=9cm]{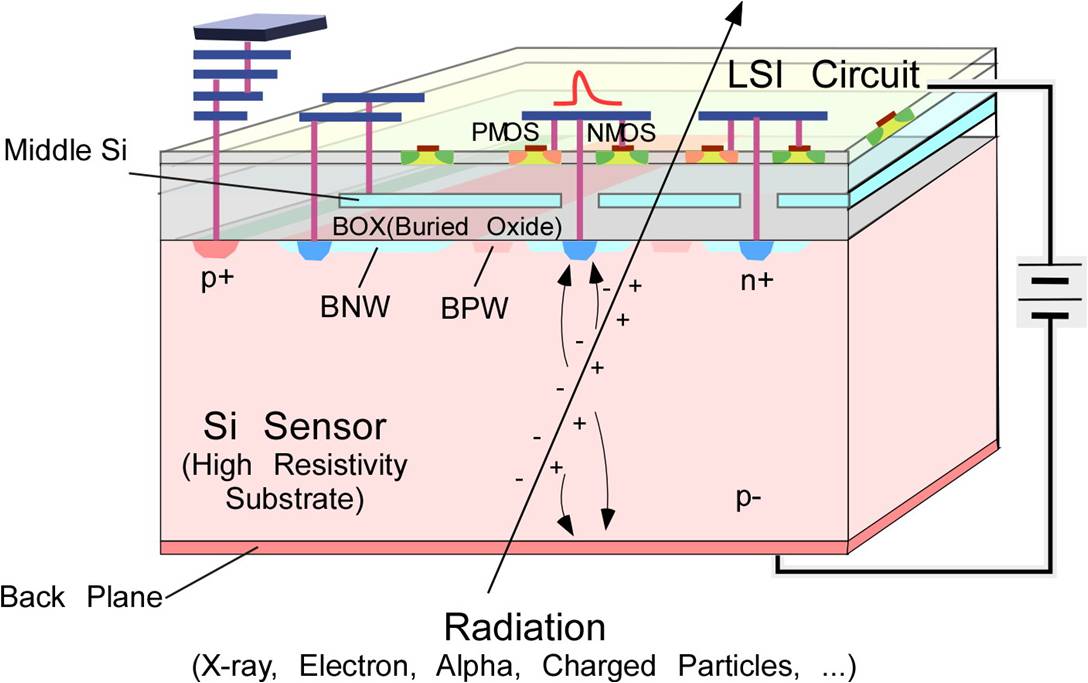}
\caption{ A scheme of cross-section of Double SOI structure }
\label{fig_dsoi}
\end{figure}		
	
	On figure \ref{fig_dsoi} the scheme of the SOI pixel structure is presented. As it was mentioned, the BOX separates the active silicon layer and the sensor area. The sensor is surrounded with BPW/BNW (Burried P-Well or N-Well) layer. The main rule of BPW/BNW is to shield the electronics from the electrical field of the sensor. 

A particle passing through the substrate ionizes the medium and the electron-hole pairs appear. The sensor is polarized, so the charges (holes in case of  the n-type sensor) flow towards the pixels and generate a signal  for the readout electronics. The main functions of the latter is to amplify, shape, sample and to transmitt off the chip the stored signals. 
Since the SOI allows to produce the readout electronics and the sensor on a one silicon wafer, thus the SOI pixel detectors are classified as the "monolithic" detectors,  opposite to  the hybrid detectors where the sensor and electronics are produced as separate chips, bump-bonded afterwards. The SOI pixels achieve similar or better results than the hybrid pixels \cite{arai1} \cite{arai2} \cite{arai3} \cite{vertex} \cite{imran1} \cite{imran2}.  The meaningful advantage of SOI detectors is its thickness, which is significantly smaller then for the hybrid detectors.  This feature makes the SOI detectors attractive for  tracking detectors applications, where  the main issue is to reduce the amount of a material  of the detector in order to decrease the Coulomb scattering.
		

In the year 2011 a new enhancement of the SOI technology appeared - the so-called "Double SOI". It consists of  a thin  conducting  $Si$ layer of silicon, set in the middle of the BOX layer volume. This additional layer  can  take over protecting functions of  the BPW/BNW layer, eliminating  a principal drawback of the BPW/BNW which is an increased capacitance of the pixel. 
The additional advantage of the Double SOI is a possible reduction of  the radiation damage effects \cite{honda}.  A  proper potential applied to the middle $Si$ layer  compensates transistor  threshold voltage shifts caused  by an electrical field originating in positive charges induced in the BOX during the irradiation. 
	
Successive approximation ADC converters have the advantage of achieving high resolution (10 bits or more) with very low power consumption \cite{culur}. Due to that the SAR ADCs are very attractive for  high conversion performance applications needed in data acquisition systems for particle detectors, spectroscopy and other fields. 

In this paper we present the current status of the SOI pixel detector development. The  first pixel detector prototype and the obtained results are presented briefly in the second section.  The third section descibes the design and architecture of new pixel detector prototypes, together with the simulation results. 
The design and simulations of a prototype 10-bit and 6-bit SAR ADC for further use in the readout electronics of the novel pixel detector are presented in section four. In section five the prototype pixel matrix, with the column ADCs integrated, is described. Finally, the conclusions are given.	



\section{First prototype of SOI pixel detector}

	A first prototype of the pixel detector was designed and fabricated in \textit{200 nm SOI Fully Depleted Low-Leakage CMOS} provided by Lapis Semiconductor Co. The principal component of the prototype detector chip was the 32 $\times$ 32 matrix of integrating type pixels, protected against the \textit{back-gate effect} by means of the Buried P-Well layer.
	
\begin{figure}[ht]
\centering
\includegraphics[trim = 0.5cm 0.5cm 0.8cm 0.5cm,clip=true, width=8.5cm]{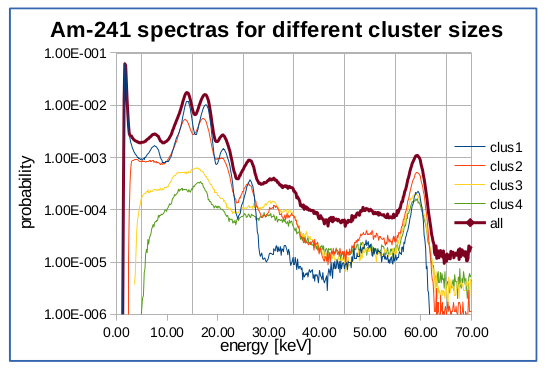}
\caption{The spectra for signal clusters of different sizes}
\label{fig_spec_kap}
\end{figure}		
	
Radiation induced charges accumulated on the small pixel capacitances generate signals high enough to be processed in a simple readout chain consisting of a source follower and a switching capacitors type  circuitry. The  latter, built as a fully differential network,  performed the Correlated Double Sampling (CDS) \cite{CDS} filtering, reducing the $kTC$, the flicker and the Fixed-Pattern noise levels.  The sensor was read out in the rolling-shutter mode, where signals from subsequent rows are transferred out without disturbing integration in other rows.
The rolling shutter is the method of data acquisition in which the image is recorded by horizontally scanning across the matrix. The advantage of this method is that the dead time of the system is shorter than in standard "snapshot" readout. Nevertheless, not all parts of the image are recorded exactly at the same time. 

The whole chip operation was synchronized by a common clock, which  frequency, and consequently the shortest integration period, were limited to  $f_{max}~=~12.5~MHz$ and  $T_{integr}~=~82~\mu s$, respectively. This limitation, originating in a speed of output amplifiers, determined the minimal achievable level of the parallel component of the Equivalent Noise Charge: 
$ENC_{par}=\frac{1}{2}(e \cdot I_{leak} \cdot T_{integr})$, where  $I_{leak}$  stays for the pixel leakage current. The total $ENC$ measured was about 115 electrons (at $60~V$ bias, $T_{integr}~=~82~\mu s$, room temp.), in which $56~e$ came from the leakage current, $34~e$ from the input transistor and the rest, i.e. 94 electrons is produced in the readout electronics (including out of chip components). 
 
The detector chips were tested with the $ ^{241}Am$ radioactive source. Figure \ref{fig_spec_kap} shows exemplary spectres obtained for signal clusters of different sizes with $T_{integr}~=~160~\mu s$; all principal energy peaks (13.9, 17.8, 20.8, 26.4 and 59.5 $keV$) plus low energy $Cu$ peak are well distinguishable. 

\section{Design of new pixel sensor}

\subsection{Source-follower based pixels and pixels with preamplifier}

\begin{figure}[h]
\centering
\includegraphics[width=8.0cm]{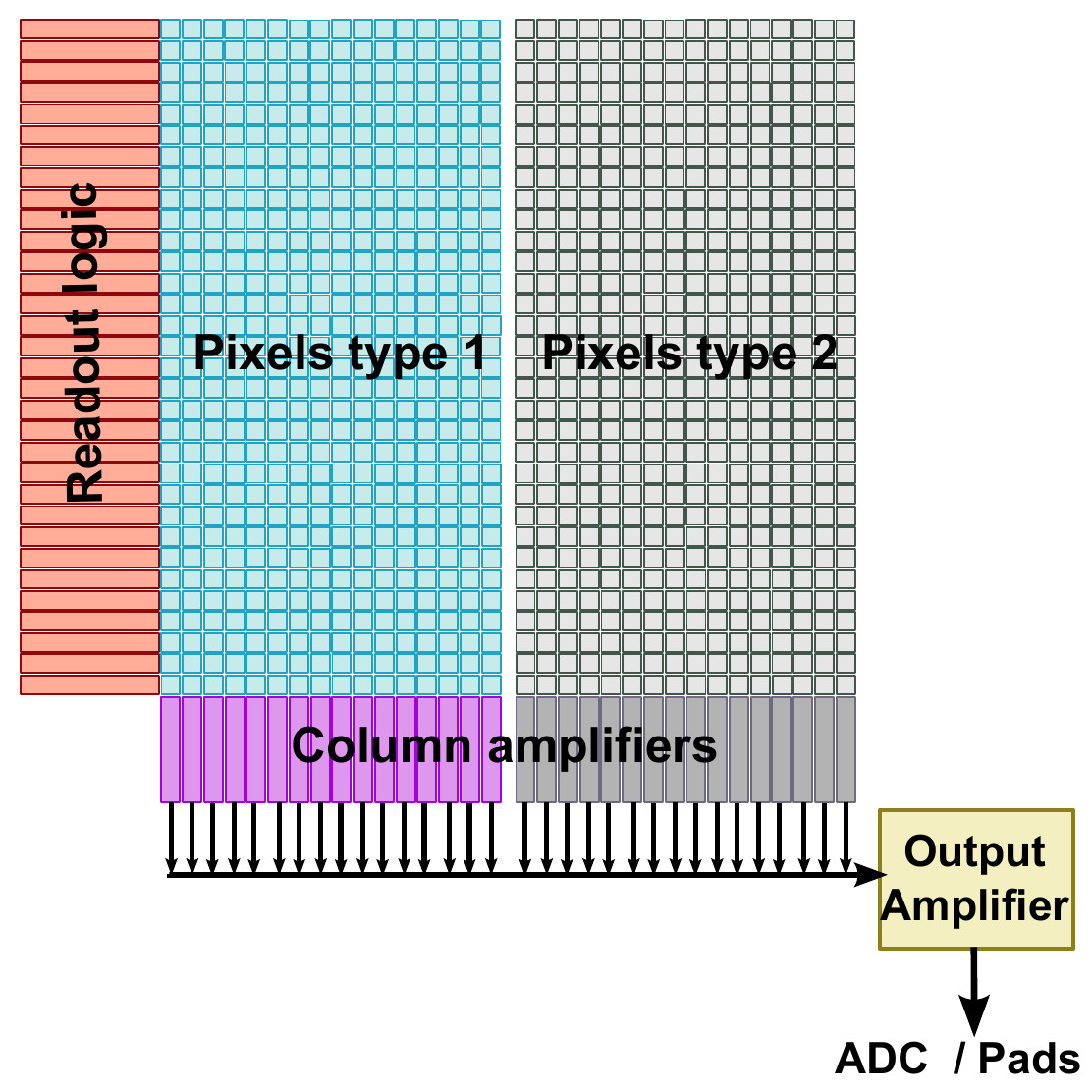}
\caption{ The block diagram of new pixel sensor }
\label{fig_design}
\end{figure}

Figure \ref{fig_design} shows the block diagram of the designed pixel sensor. It consists of two different pixels matrices, the readout logic and the column and output amplifiers. Each matrix contains 36 rows and 8 columns of pixels 30 $\times$ 30 $\mu m$ each. The Double SOI feature (additional $Si$ layer) was used beneath each pixel in the matrix.

The readout logics works in the \textit{rolling shutter} mode. 
\begin{figure}[ht]
\centering
\includegraphics[width=9cm]{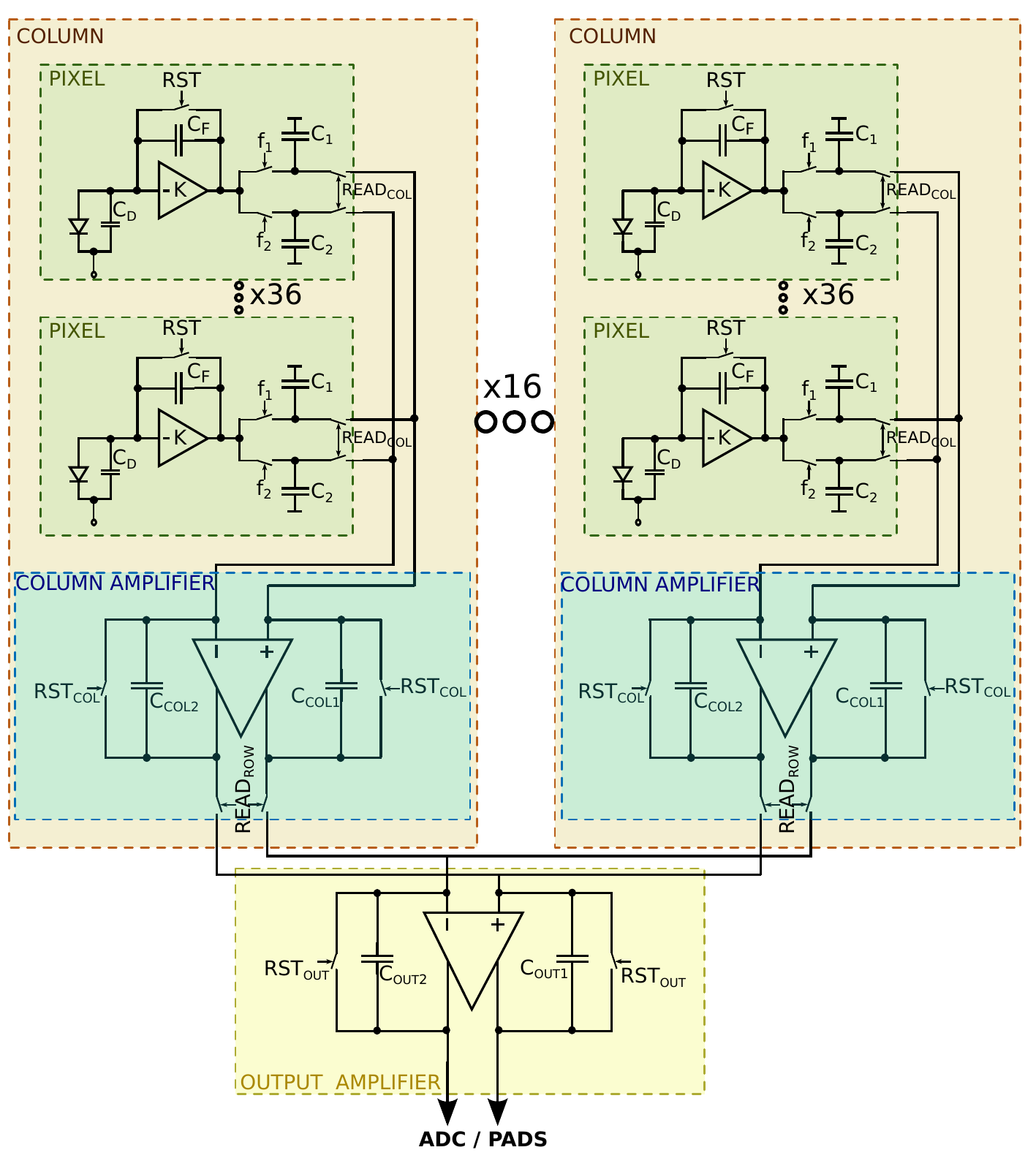}
\caption{The scheme of the designed front-end electronics.}
\label{fig_design2}
\end{figure}
In figure \ref{fig_design2} the schematic diagrams of the designed front-end subcircuits are presented. As it is shown, the single column consists of a set of 36 pixels which outputs are connected to the column amplifier. The pixel matrix readout is controlled by two signals: $READ_{ROW}$ and $READ_{COL}$. In the first step the $READ_{ROW}$ selects a single row to be read. In practice the output capacitances of the pixel ($C_1$ and $C_2$) are connected to the inputs of the columns amplifiers. In the next phase the $READ_{COL}$ is responsible for connecting sequentially each column amplifier to the output amplifier. The information from output amplifier is finally sent to an analog to digital converter (ADC) or to output pads. 

\subsubsection{Source-follower based pixels}
One of the matrices (Pixels type 2 in figure \ref{fig_design}) uses a simple source-follower based pixel readout. The pixel is very similar to the first prototype described in previous section, although the design was optimized with regarding to signal to noise ratio.  Both the pixel sensor and the source-follower front-end are slightly modified. 
	
\subsubsection{Pixels with preamplifier}
The second matrix (Pixels type 1 in figure \ref{fig_design}) uses a new preamplifier-based front-end.
To study the effect of the Dpuble SOI feature half of the pixels in this matrix were drawn using standard BPW/BNW layer and other half without it.	

\begin{multicols}{2}
\begin{figure}[H]
\centering
\includegraphics[width=0.5\textwidth]{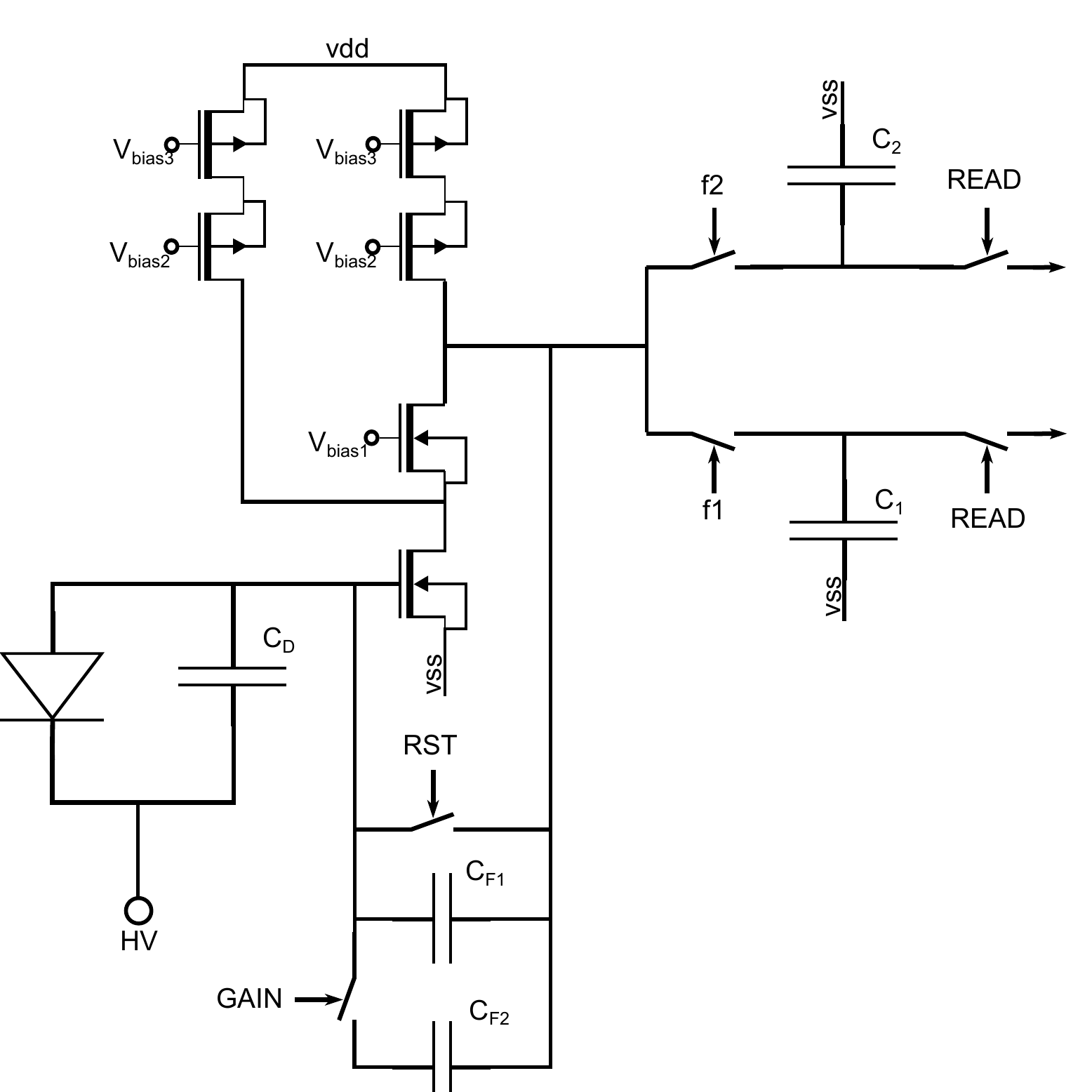}
\caption{ A scheme of the pre-amplifier circuit }
\label{fig_preamp}
\end{figure}
\begin{figure}[H]
\centering
\includegraphics[width=0.5\textwidth]{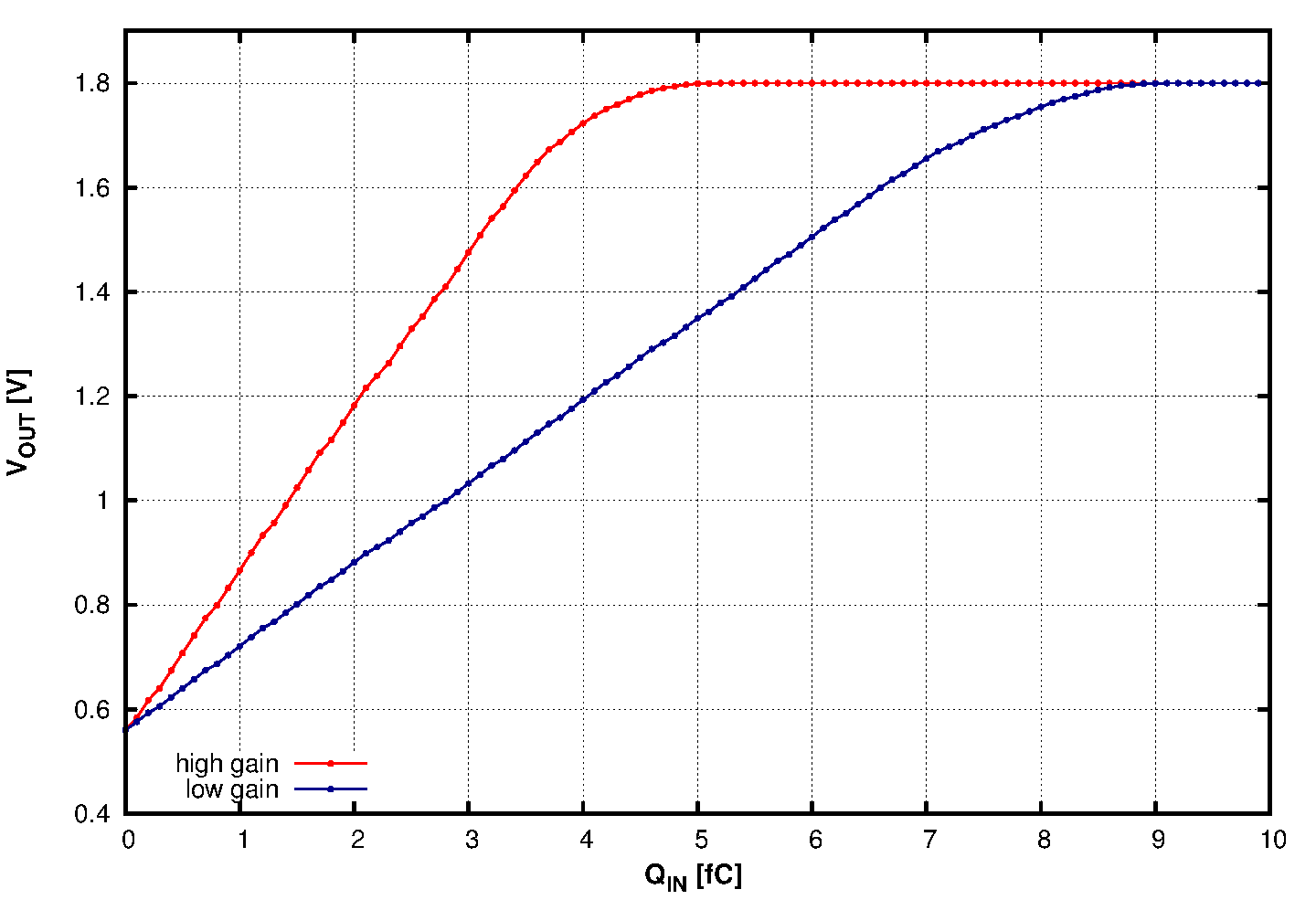}
\caption{The pre-amplifier output voltage in function of collected charge }
\label{fig_wyk1}
\end{figure}	

\end{multicols}

One of the most  significant enhancements, in comparison to the first prototype, is the newly designed pre-amplifier. The main rule of the pre-amplifier is to convert the charge collected by the sensor into the voltage information. Figure \ref{fig_preamp} shows the scheme of the designed pre-amplifier circuit. 
The pre-amplifier is designed as a telescope cascode with an additional current source for higher output resistance. The extra current flows through the input transistor to increase its transconductance. The pre-amplifier is designed without resistive feedback loop. It allows to set two different gains by selecting appropriate feedback capacitance $C_{F2}$. 
In order to eliminate the undesired offset of the input stage, to reduce the noise and to obtain the differential signal, the correlated double sampling (CDS) is used \cite{CDS}. The CDS method is based on two measurements performed in a short time interval. According to figure \ref{fig_preamp}, in the first phase the $f1$ switch is conducting and the $f2$ is switched off, therefore the baseline is sampled on the $C_1$ capacitance. Next, the $f1$ is switched off and $f2$ switch is conducting. In that case, the voltage which is a sum of the baseline voltage and the voltage corresponding to charge collected by the sensor, is sampled on the $C_2$ capacitance. On figure \ref{fig_wyk1} the plot of pre-amplifier output voltage in function of collected charge, for two different gains, is presented.

\begin{table*}[ht]
\caption{The noise analysis summary}\label{tab}
\centering
\resizebox{\textwidth}{!}{
\begin{tabular}{|l|c|c|c|c|c|}
\hline
\textbf{SETTNGS} & \textbf{ENC [$e^-$]} &  \textbf{ENC [$e^-$]}  & \textbf{Noise RMS [$mV$]} &  \textbf{Noise RMS [$mV$]} & Gain [$\frac{ \mu V}{e^-}$] \\ 
                 &       &  \textbf{using CDS}  &     &  \textbf{using CDS} & \\ 
\hline\hline

Low gain                  & 45 & 42 & 0.93 & 0.88 & 20.3 \\ 
Low gain with shot noise  & 226 & 43 & 4.6 & 0.9 & 20.3 \\ 
High gain                 & 36 & 33.7 & 1.4 & 1.3 & 39.4 \\ 
High gain with shot noise & 163 & 34 & 6.3 & 1.4 & 39.4 \\ 

\hline
\end{tabular} }
\end{table*}

Since the pixel leakage current is not known apriori, for each gain the results are presented without and with the sensor shot noise. The shot noise was estimated from the measurements done with the first prototype, where the leakage current was on the level of $2$~$pA$ per pixel. Table~\ref{tab} presents the noise analysis summary for both gains and for the readout performed without and with the CDS technique. 

\subsubsection{Column amplifier}

\begin{figure*}[ht]
\centering
\includegraphics[width=0.99\textwidth]{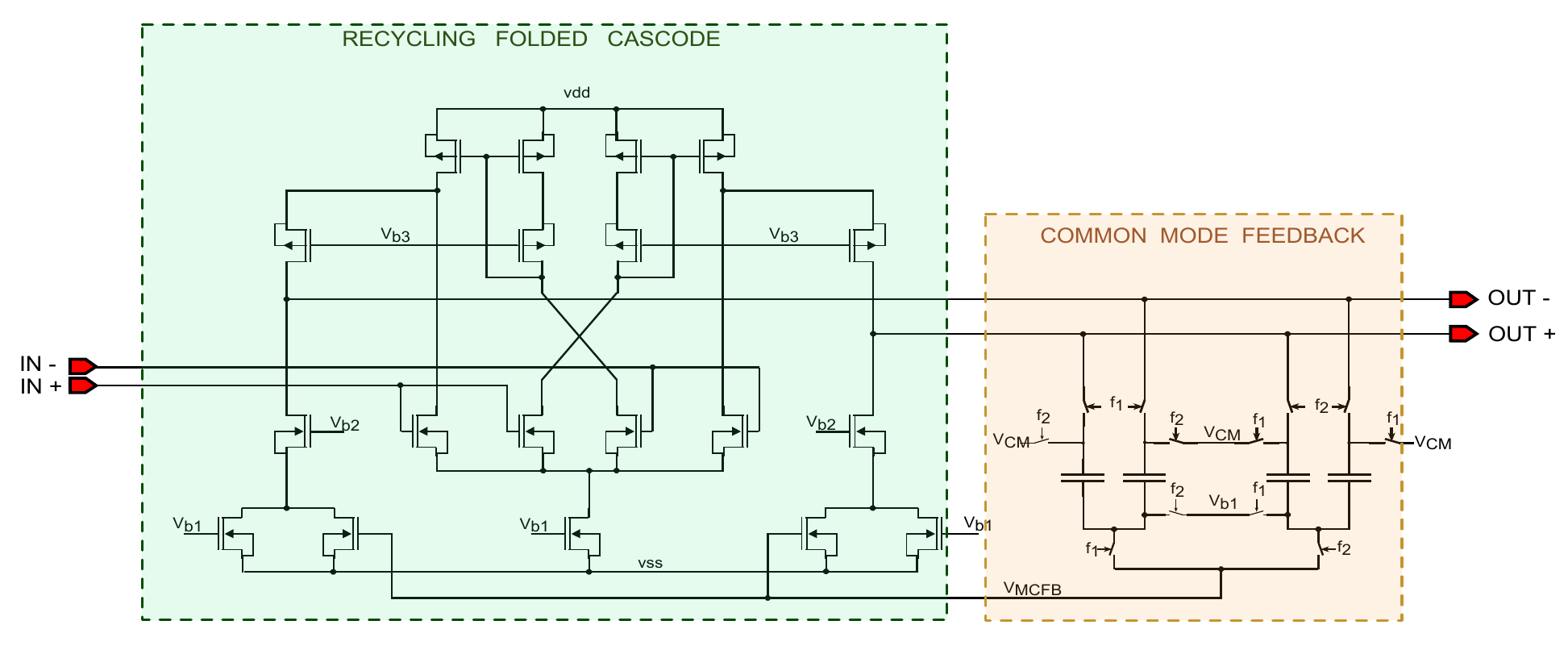}
\caption{ A scheme of a designed column amplifier }
\label{fig_colamp}
\end{figure*}

The overriding goal in the column amplifier design is to minimize its power consumption. Due to that the recycling folded cascode~\cite{RFC} with capacitive common mode feedback is chosen. This architecture is presented on figure \ref{fig_colamp}. The main parameters of the described output amplifier are: 
	\begin{itemize}
	\item phase matgin - $68^o$
	\item gain - $750~\frac{V}{V}$
	\item GBW - $170~MHz$
	\item power consumption - $72~\mu W$
	\end{itemize}

\subsection{Self-triggering pixels}
A third, more complex, self-triggering type of pixel sensor was designed in view of possible time and amplitude measurements. A small prototype matrix contains 8 rows and 4 columns of pixels 30 $\times$ 100 $\mu m$ each. The Double SOI feature (additional $Si$ layer) was used beneath each pixel in the matrix.

\begin{figure}[ht]
\centering
\includegraphics[width=0.7\textwidth]{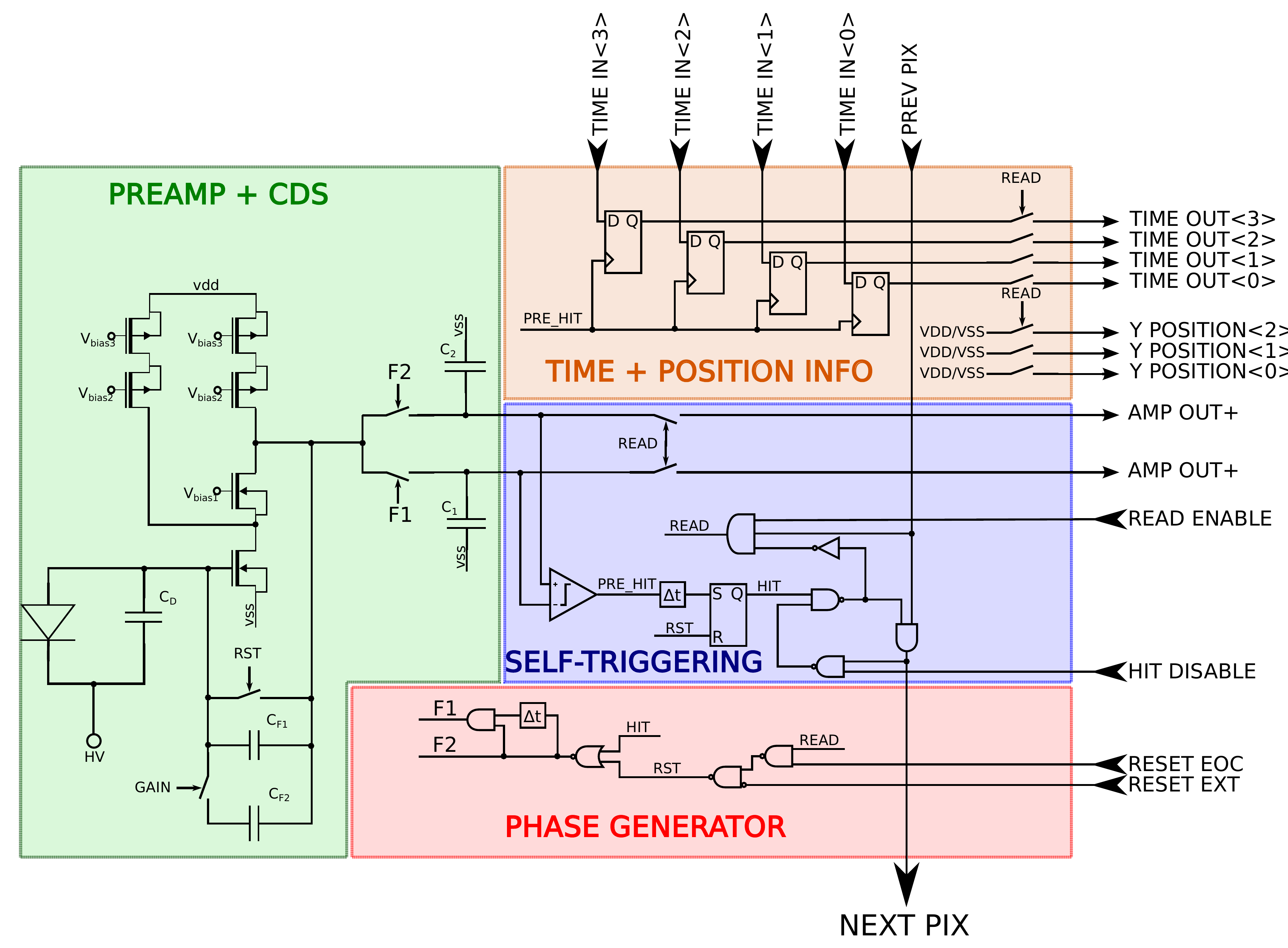}
\caption{The scheme of self-triggering pixel.}
\label{self_trig_pix}
\end{figure}

The scheme block of designed self-triggering pixel is presented on figure \ref{self_trig_pix}. The analog part of this pixel (pre-amplifier and CDS) are mostly the same as in the previously described preamplifier based pixel. The new subcircuit are: a simple logic for self-triggering, a discriminator and four flip-flops that store time of hit arrival. There is also another logic block providing proper signals for the CDS called the phase generator. 

The idea of self-triggering pixel was to find solution dedicated for low-occupancy issues, which are familiar for tracking systems. The standard snapshot or rolling shutter readouts are ineffective, when approximately 99\% of read out pixels do not containt any information. The general idea of self-triggering is to provide a "READ" signal for the pixel which was hit and disable all other pixels. The pixel with hit is read out and then the other one if there is anything to be read. This algorithm is provided by the discriminator and the self-triggering logic in the proposed pixel. 

\begin{figure}[ht]
\centering
\includegraphics[width=0.7\textwidth]{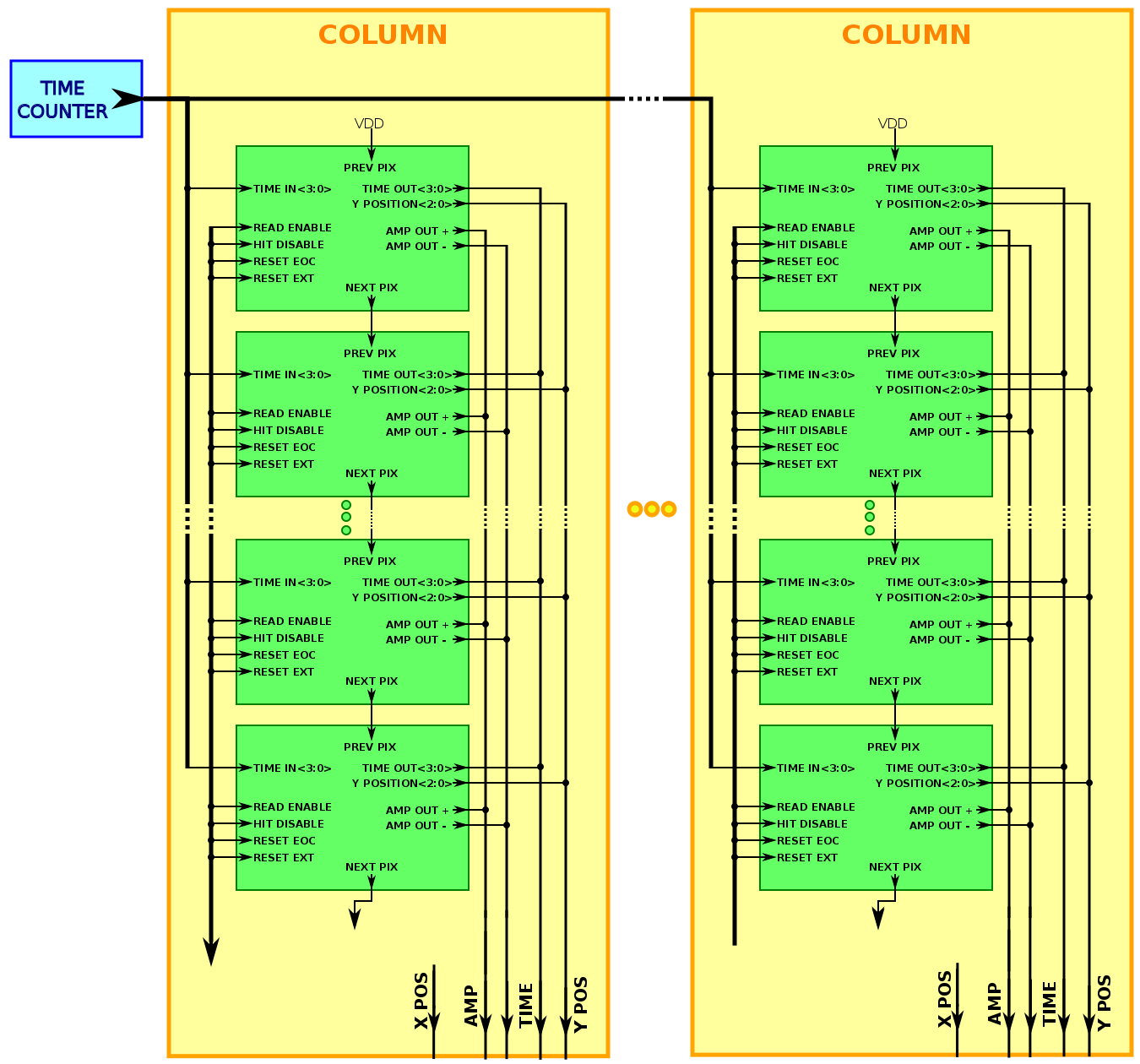}
\caption{The scheme of the self-triggering pixel matrix.}
\label{self_trig_column}
\end{figure}

In figure \ref{self_trig_column} the scheme of test matrix is shown. It is only a small prototype, dedicated for tests and debugging of the self-triggering system. The matrix is 8$\times$4 pixel large and each column is fully independent, so in the future it could be easily enlarged to the needed size. Outside the matrix there is a 4b time counter, that provides information for whole matrix. 

\section{Design of 10-bit and 6-bit SAR ADC}
\begin{figure}[ht]
\centering
\includegraphics[width=0.7\textwidth]{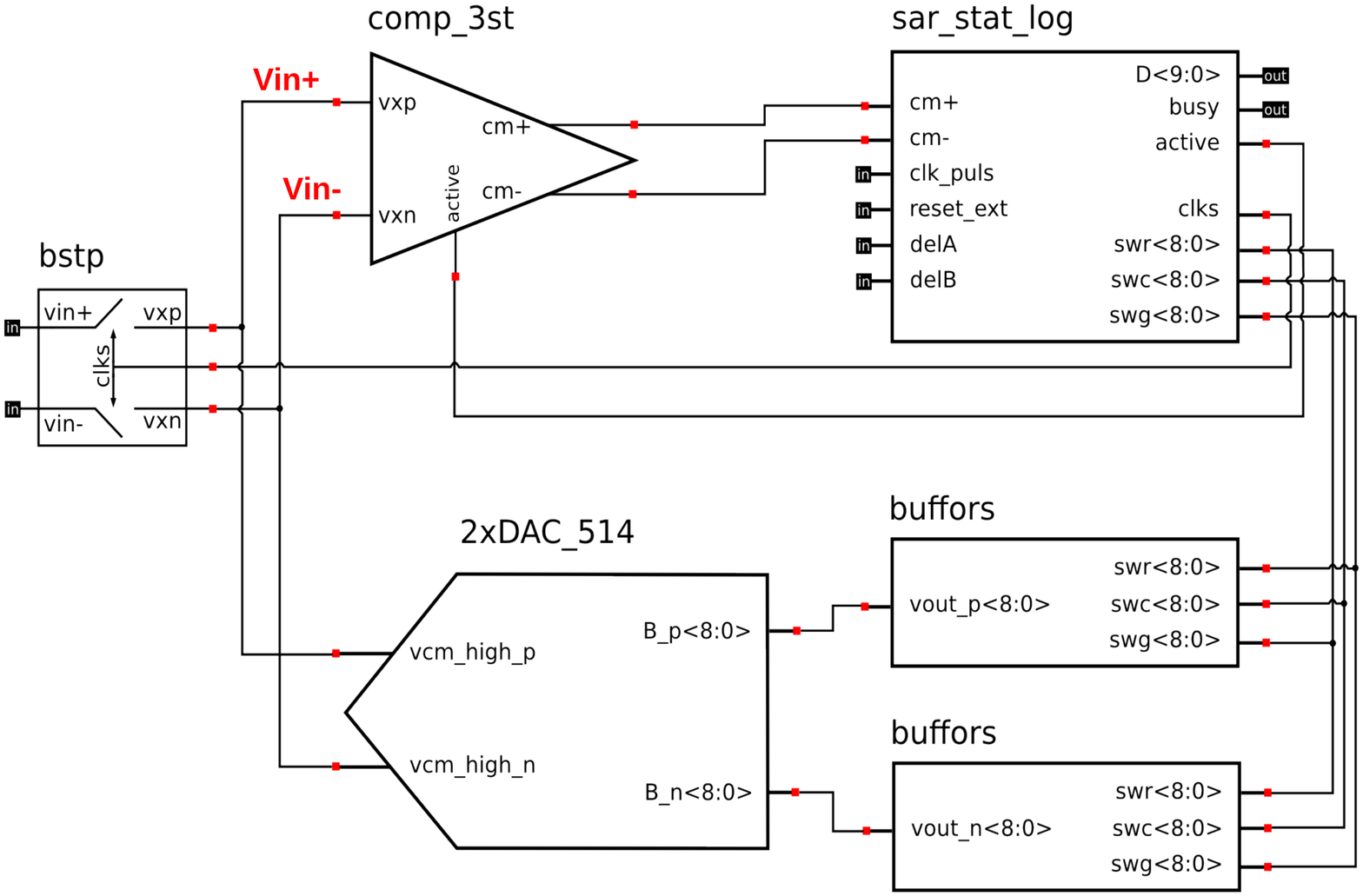}
\caption{The scheme of designed ADC}
\label{saradc}
\end{figure}

The analog-to-digital conversion would be a very requested feature of the readout ASIC. The main difficulty in implementation of the ADC is its excessive power consumption.
The Successive Approximation Register Analog-to-Digital Converter (SAR ADC) is well known for its power efficiency. This feature, together with other advantages of modern sub-micron technologies, makes the SAR ADC a perfect candidate for modern readout ASICs. We are developing two ADC prototypes: a 10-bit SAR ADC for precise amplitude measurements and a lower resolution 6-bit SAR ADC for tracking or similar applications. The design of the 10-bit ADC is described in the following. The 6-bit one was designed using the same architecture and the same (or slightly modified) blocks.
In figure~\ref{saradc} the architecture of the designed SAR ADC is presented. It consists of four basic subcircuits:
\begin{itemize}
\item \textbf{bootstrapped sampling switch} - to sample the input voltage ($V_{in}$). The design of this circuit is focused on increased linearity of the ADC. In order to decrease the sampling resistance and make it independent on signal amplitude a bootstrapped MOS switches~\cite{bootstrap_switch} are implemented in the differential S/H input. 
\item \textbf{digital-to-analog converter (DAC)} - to generate and hold the voltages which are compared with input voltages $V_{in+}$ and $V_{in-}$. 
\item \textbf{comparator} - to compare the differential outputs of DACs with $V_{in+}$ and $V_{in-}$. A fully dynamic 3-stage comparator ~\cite{comp}~was implemented. Withe a dynamic comparator the  designed ADC does not have any block drawing static currents and so the ADC has zero static power consumption.
\item \textbf{control logic} - to control the operation of the entire converter. The presented design uses asynchronous logic, which provides significant power saving (clock tree is not needed). The whole logic is designed using static elements to provide more predictable behaviour. 
\end{itemize}
In addition to the above mentioned functional blocks the buffers are also needed for fast charging of DAC capacitance array. Fully differential implementation (two DACs, two bootstrapped switches and two buffers) improves linearity of the circuit and  overall resistance to disturbances.

\subsection{Capacitive DAC}

\begin{figure}[h]
\centering
\includegraphics[width=0.7\textwidth]{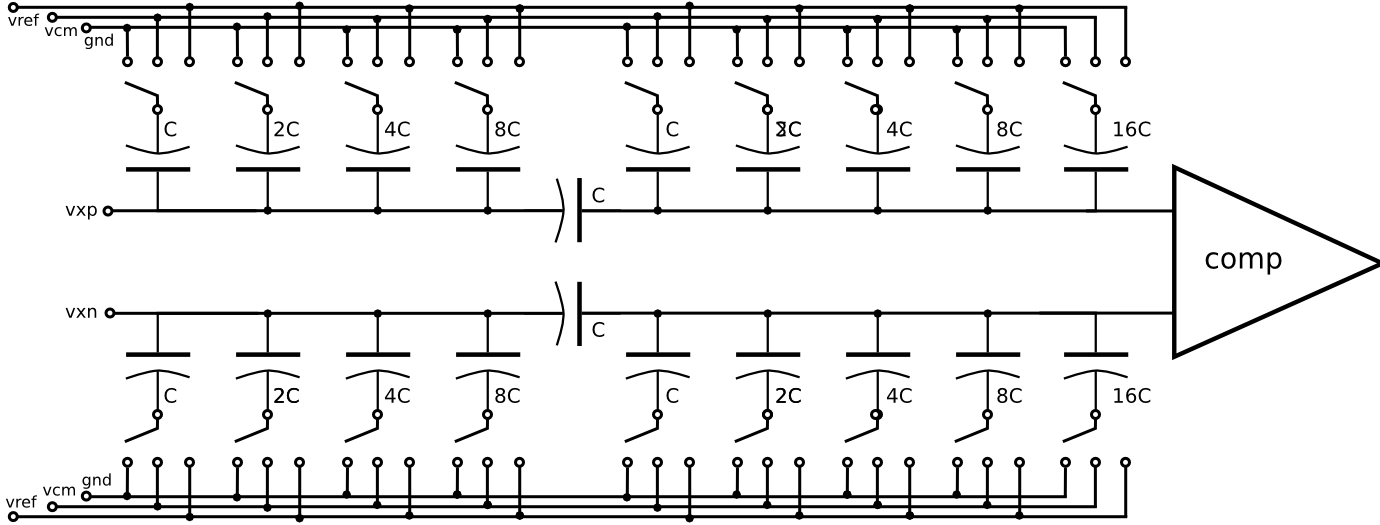}
\caption{The scheme of designed DAC}
\label{dac}
\end{figure}

The DAC converter is one of the most sensitive parts of the ADC. The differential, segmented DAC with merge capacitor switching (MCS) scheme was implemented~\cite{mcs}. It allows to achieve ultra low power switching performance. The MCS scheme achieves $93.4\%$ less switching energy as compared to the conventional SAR architecture.
The scheme of the DAC is presented in figure \ref{dac}.

The main advantage of splitting the DAC array is a possibility to bypass the problem of relatively high minimal capacitance of $mimcap$ available in the used SOI CMOS technology. In  the proposed design the single capacitance in DAC array is 39~fF. The DAC segmentation decreases the effective capacitance and due to that lowers the power consumption and reduces the circuits area.

\subsection{Simulation results}

In the table \ref{tab4} the simulated performance results of designed 10-bit and 6-bit ADCs are shown. The parameters with $\bullet$ were not simulated yet, because of convergance problems. 

\begin{table}[ht]
\centering
\caption{The simulated performance of 10-bit and 6-bit SAR ADC}
\vspace{0.1cm}
\label{tab4}
\begin{tabular}{|c|c|c|}

\hline
\textbf{PARAMETER} & \textbf{10b ADC} & \textbf{6b ADC} \\
\hline
Supply voltage & \multicolumn{2}{|c|}{1.8V} \\
\hline
Power consumption & $\sim$~900~$\mu W~at~10 MHz$ & $\sim$~650~$\mu W~at~20MHz$ \\
\hline
Max $f_{sample}$ (schematic) & 20 MHz & 100 MHz\\
\hline
Max $f_{sample}$  (post-layout) & $\bullet$ & 20 MHz\\
\hline
Input Capacitance & 2pF & 310fF \\
\hline
ENOB (schematic) & 9.95 & 6.0 \\
\hline
ENOB (post-layout) & 9.5 (with DAC, bstp and comp) & 5.98 \\
\hline
ADC area & 310 $\mu$m $\times$ 190 $\mu$m & 300 $\mu$m $\times$ 50 $\mu$m\\
\hline
\end{tabular}
\end{table}

\section{Self-triggering pixel matrix with column ADC readout}

\begin{figure}[H]
\centering
\includegraphics[width=0.5\textwidth]{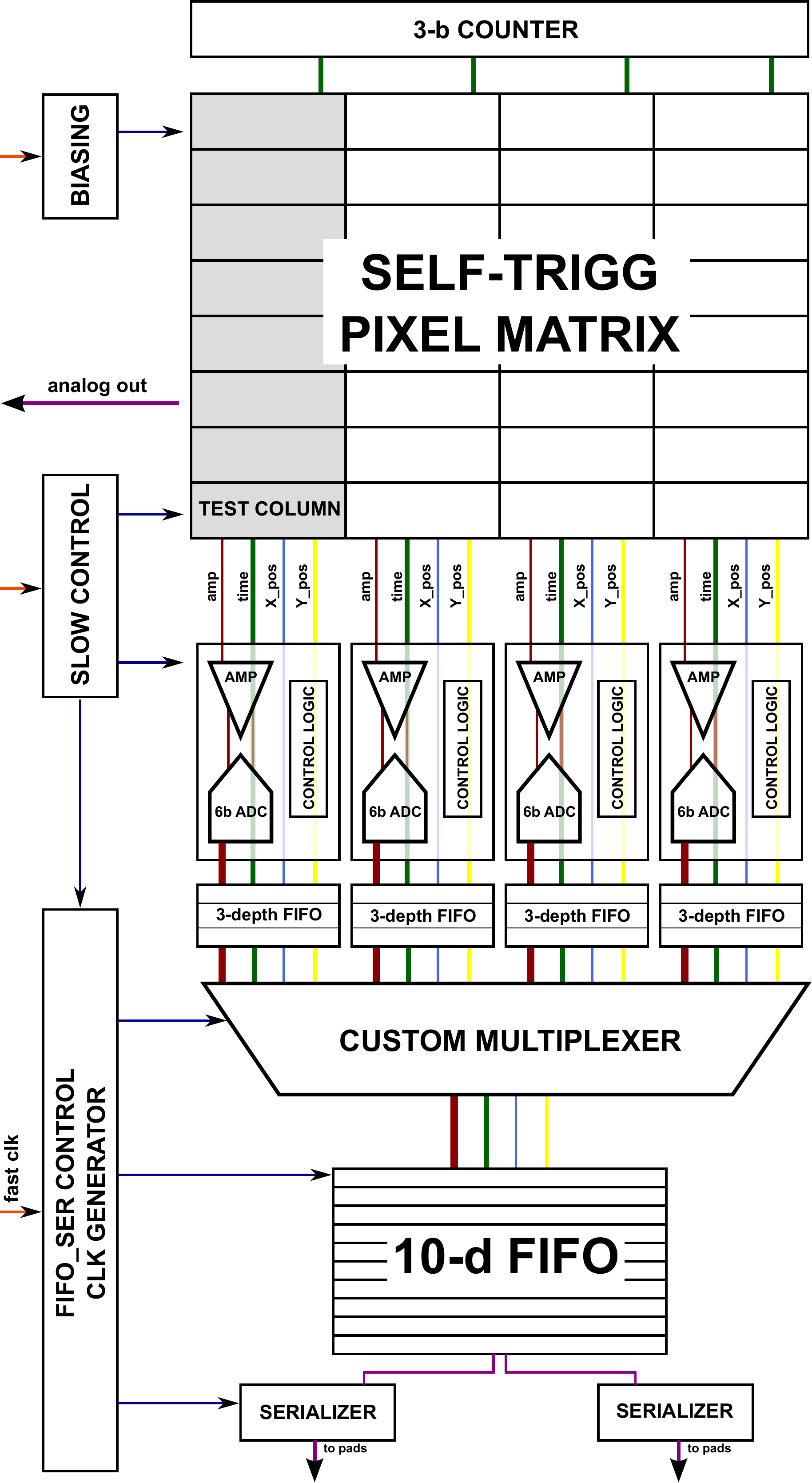}
\caption{Block diagram of the self-triggering pixel matrix with column ADC readout}
\label{fig_full_readout}
\end{figure}

In the previous sections the self-triggering pixels and ADC were described. These key blocks were integrated, together with other digital blocks, into a small but relatively complex pixel detector readout system shown in figure~\ref{fig_full_readout}. The self-triggering matrix has 4 columns each 8 pixel. One 4-bit counter provides the time information for pixels. From each column the following information package is sent: amplitude, 4-bit time, and position information. The amplitude signal is amplified, digitalized and all the data from one column are stored in 3-depth FIFO. After that, there is a block that works as a data selector: it chooses columns one by one and controls the data transfer to one big FIFO, containing the data from whole matrix. Finally, the data are sent to two serializators (10 bits per each) and are send out through a differential LVDS interface.  
The whole system needs a set of control bits, that are provided by Slow Control. Also the fast
external clock is needed. The readout speed of whole system was simulated up to 200MHz. Moreover, there is a one test column with analog pixel output, that allows to control the pixel performace directly from the matrix.

\section{Conclusion}
In this paper, the pixel and the readout electronics designed in \textit{200 $nm$ Silicon-On-Insulator Fully-Depleted Low-Leakage CMOS}, offered by Lapis Semiconductor Co., is presented and discussed. The new pixels use few versions of newly designed front-end electronics: the optimized source follower front-end; the charge amplifier configuration, in order to obtain the lowest noise; and the self-triggering front-end, for the most efficient readout, together with time and amplitude measurements. A small matrices were designed for each kind of pixels.  The Double SOI  feature is included in the design and will be studied  as a solution against the radiation damage effects. Two versions of low-power SAR ADC were also designed and described. A 10-bit SAR ADC for precise amplitude measurements and a 6-bit version for tracking applications. 
All designed blocks were integrated in a prototype ASIC and submitted to fabrication.

\Acknowledgements
This work was supported by the Polish National Science Centre (NCN) under grant no. 2012/07/B/ST2/03752.

\end{document}

%% file: econfmacros.tex
\def\beq{\begin{equation}}
\def\eeq#1{\label{#1}\end{equation}}
\def\eeqn{\end{equation}}

\def\beqa{\begin{eqnarray}}
\def\eeqa#1{\label{#1}\end{eqnarray}}
\def\eeqan{\end{eqnarray}}

\let\bar=\overbar

\def\Dslash{\not{\hbox{\kern-4pt $D$}}}
\def\dslash{\not{\hbox{\kern-2pt $\del$}}}

\def\msb{{\bar{\ssstyle M \kern -1pt S}}}

